\documentstyle[12pt]{article}

\begin{document}

\title{Dynamic Magnetization-Reversal Transition in the Ising Model}
\author{Arkajyoti Misra and Bikas K. Chakrabarti \\
Saha Institute of Nuclear Physics\\
1/AF Bidhannangar\\
Calcutta 700 064\\
India.}
\maketitle

\begin{abstract}

We report the results of mean field and the Monte Carlo study of
the dynamic magnetization-reversal transition in the Ising model, 
brought about by the application of an external
field pulse applied in opposition to the existing order before the application 
of the pulse. 
The transition occurs
at a temperature $T$ below the static critical temperature $T_c$ without any external field.
The transition occurs when the system, perturbed by the external field pulse competing with 
the existing order,  jumps 
from one minimum of free energy to the other
after the withdrawal of the pulse. 
The parameters controlling the
transition are the strength $h_p$ and the duration $\Delta t$ of the pulse.
In the mean field case, approximate analytical expression is obtained for
the phase boundary which agrees well with that obtained numerically
in the small $\Delta t$ and large $T$ limit.
The order parameter of the transition has been identified and is observed to vary
continuously near the transition.
The order parameter exponent $\beta$ was estimated both for the mean field ($\beta =1$) 
and the Monte Carlo ($\beta = 0.90 \pm 0.02$ in two dimension) cases. The transition shows 
a "critical slowing-down" type behaviour near the phase
boundary with diverging relaxation time. The divergence was found to be logarithmic
in the mean field case and exponential in the Monte Carlo case. The finite size 
scaling technique was employed to estimate the correlation length exponent $\nu$ 
($= 1.5 \pm 0.3$ in two dimension) in the Monte Carlo case.

\end{abstract}


\section{Introduction}

The dynamic response of pure Ising systems to time dependent magnetic 
fields is being studied intensively these days (see e.g. \cite{ac,rik1} and references 
therein). In particular, the response of Ising systems to pulsed fields have recently 
been investigated \cite{abc,mc}. The pulse can be either 
"positive" or "negative". At temperatures $T$
below the critical temperature $T_c$ of the 
corresponding static case (without any external field), majority of the 
spins orient themselves
to a particular direction giving rise to the prevalent order. 
If the external field pulse is applied along the direction of the 
existing order, it is called a positive pulse and if the pulse is applied 
opposite to the 
existing order, it is called a negative pulse. The effect of positive pulse has
been studied by Acharyya et al \cite{abc}, whereas the occurrence of a
magnetization-reversal transition as a result of the application of the negative
pulse has been reported in an earlier work \cite{mc}. We report here the 
results of detailed investigation of this dynamic magnetization-reversal 
transition for pure Ising models under a pulsed field.

In the absence of any symmetry breaking field, for temperatures below the critical 
temperature of the corresponding
static case ($T < T_c$), there are two equivalent free energy minima with
 average magnetizations 
$+m_0$ and $-m_0$. If in the ordered state the equilibrium magnetization is 
$+m_0$ (say) and the pulse is applied in the direction 
opposite to the existing order, then temporarily during the pulse
period the free energy minimum with magnetization $-m_0$ will be brought down 
compared to that with $+m_0$. If this asymmetry is made permanent, then 
any nonzero field (strength), which is responsible for the asymmetry,  would eventually induce 
a transition from $+m_0$ to $-m_0$. Instead, if the field is applied 
in the form of a pulse, the asymmetry in the free energy wells is removed after a finite
period of time. In that case, the point of interest lies in the combination of 
the pulse height or strength ($h_p$) and its width or duration ($\Delta t$) that can 
give rise to the transition from $+m_0$ to $-m_0$. We call this a 
magnetization-reversal transition. A crucial point about the 
transition is that it is not necessary that the system attains its final 
equilibrium
magnetization $-m_0$ during the presence of the pulse; the combination of $h_p$
and $\Delta t$ should be such that the final equilibrium state is attained at any
subsequent time, even long time after the pulse is withdrawn. The phase boundary,
giving the minimal combination of $h_p$ and $\Delta t$ necessary for the transition,
depends on the temperature. As $T \rightarrow T_c$, the magnetization reversal 
transition
occurs at lower values of $h_p$ and/or $\Delta t$ and the transition disappears at
$T \ge T_c$.

In this paper we have given both the mean field (MF) and the Monte Carlo (MC) 
results for the transition. The MC studies have been carried out for a two
dimensional ($d=2$) lattice of Ising spins. The phase boundaries in the $h_p-\Delta 
t$ plane (each for a fixed $T$) are obtained for both the MF and the MC cases. 
Approximate analytical expressions are also obtained and compared to these phase 
boundaries.
The order parameter (${\cal O}$) for the dynamic transition has been identified and 
at a fixed $T$ its variation with the driving 
parameters $h_p$ and $\Delta t$ has been studied. The observed continuous 
variation of ${\cal O}$ indicates the nature of the transition to be comparable to the 
second order type static transitions. The critical exponents for the order
parameter variations near the phase boundary has been obtained for both the MF
and MC cases. We also employed the finite size scaling method (see e.g 
\cite{barber}) to estimate the correlation length exponent $\nu$ 
for the transition  in the MC studies. We observe 
significant "critical slowing down" near the phase boundary and the behaviour of
the relaxation time $\tau$ has been studied in both the cases. In the MF case, we
have obtained an approximate analytical expression for $\tau$ indicating clearly
 a different kind of divergence of $\tau$ as compared to that in the static case.

\section {Model}

We have taken Ising Model for both the numerical simulation and mean
field study. The Hamiltonian of nearest neighbour Ising system without
any disorder is

\begin{equation}
 H=-J \sum _{<ij>}S_{i}S_{j}-h(t)\sum _{i}S_{i} ,
\end{equation}

\noindent
where \( S_{i}=\pm 1 \) represents the Ising spins at lattice site \( i \) 
and \( J \) denotes the
nearest neighbour interaction strength. The time dependent external
magnetic field \( h(t) \) is applied in the form of a pulse of duration \( \Delta t \) 

\begin{eqnarray}
h(t) & = - h_p , & {\rm for} ~~~t_0 < t < t_0+\Delta t \nonumber \\
     & = 0 , &  {\rm otherwise} .
\label{eq:stepfn}
\end{eqnarray}

\noindent
$t_0$ is taken to be much larger than the relaxation time of the 
unperturbed system so that the system is guaranteed to reach a state
of equilibrium before the pulse is applied. The average magnetization
$m$ is given by $<S_i>$, where the angular brackets represent thermal average.
By the time $t=t_0$, the system has reached its equilibrium state with the 
magnetization $m(t)=\pm m_0$.  This is the state before the application of the pulse, 
which is applied at time $t=t_0$ to compete with the initial magnetization.
We then want to look at the dynamics of the system under a pulsed
field starting with the initial condition $m(t_0) = +m_0$ (say).
Depending on the strength (\( h_{p} \)) or the duration (\( \Delta t \)) 
of the pulse,
the system has two choices after the pulse is withdrawn: it can either
go back to the original ordered state (\(m(t=\infty) =  +m_{0} \)), or it can 
switch to the
other equivalent equilibrium ordered state (\(m(t=\infty) = -m_{0} \)). Fig. 1 
shows schematically these behaviours. The result, at any finite temperature below 
$T_c$, naturally
depends on the strength and the duration of the pulse. Specifically, we observe 
that 
the result depends on $m_w \equiv m(t_0+\Delta t)$, the average magnetization 
at the time of withdrawal of the field.
The sign of $m_w$, which in turn depends on the 
combination of $h_p$ and $\Delta t$, governs the transition. $\mid m_w \mid$ was
found out to be the appropriate candidate for the order parameter: if $m_w$ becomes 
negative, on an average one observes a magnetization-reversal transition; 
however if $m_w$ is positive, the system is most likely to return back to its
original equilibrium state. In
fact, it is guaranteed to be so in the mean field case. However in MC simulations,
there exists occasional fluctuations due to which the system may finally arrive
at the $+m_0$ state starting with negative $m_w$ and vice versa.
We have identified the $h_p-\Delta t$ phase boundary at a particular
temperature which gives the optimum combination of the driving parameters
($h_p$ and $\Delta t$) that can force the system to a final state with magnetization 
$-m_0$, starting from an initial state with magnetization $+m_0$ or vice versa.
We define the relaxation
time $\tau$ as the time taken by the system to reach its final equilibrium
state from the time of withdrawal of the field (at $t = t_0+\Delta t$).
The relaxation time  increases as the value of $\mid m_w \mid$ approaches
 zero or equivalently as one approaches the phase boundary from either side.
In Fig. 2 the typical MC results (on a square lattice) show how the transition 
can be brought about by either increasing $h_p$ (see Fig. 2(a)) or $\Delta t$ 
(see Fig. 2(b)). One can also note from these figures how the relaxation time 
$\tau$ increases as one approaches the phase boundary.

\section{Mean Field Study}

The mean field equation of motion for the average magnetization $m(t)$ of the 
system is

\begin{equation}
 \frac{dm}{dt}=-m+{\rm tanh} \left( \frac{m+h(t)}{T} \right) ,
\label{eq:mfeq}
\end{equation}

\noindent
where \( h(t) \)is given by (\ref{eq:stepfn}). 
Here we have assumed $Jn=1$, where $n$ is the lattice coordination number. 
With this choice,
the critical temperature in the static limit ($T_c$) becomes unity. All the 
mean field calculations are performed, therefore, at a temperature $T<1$. 
The equation
was solved numerically to obtain the phase boundaries, shown in Fig. 3. 
A point on a particular phase boundary gives the optimal combination
of \( h_{p} \) and \( \Delta t \) that can induce the transition from a state 
with magnetization $+m_0$ to $-m_0$, where $m_0=\tanh(m_0/T)$. The axes side
of the boundaries correspond to the return to original equilibrium state,
whereas one gets
a magnetization-reversal transition for combinations of $h_p$ and $\Delta t$ beyond 
the phase boundary. Because of the absence of any
fluctuations in the mean field equation (\ref{eq:mfeq}), there exists a finite
coercive field for $T<T_c$ and therefore one cannot 
bring about the transition just by increasing $\Delta t$ if $h_{p}$ does not 
exceed the coercive
field value. Hence the phase boundary becomes parallel to \( \Delta t \) axis 
for large values of \( \Delta t \).

For large $T$, the mean field phase boundary can be estimated
approximately by solving the linearized mean field equation
\begin{equation}
\frac{dm}{dt} = -\epsilon m+\frac{h(t)}{T}~;~~ \epsilon=\frac{1-T}{T},
\label{eq:linear}
\end{equation}
where $h(t)$ is given by (\ref{eq:stepfn}). The solution of 
(\ref{eq:linear}) for $t > t_0$ can be written as 
\begin{equation}
m(t) = \left (m_0 - \frac{h_p}{\epsilon T} \right ) \exp[\epsilon (t-t_0)]
+ \frac{h_p}{\epsilon T}.
\label{eq:mfsol}
\end{equation}
It may be noted from the solution (\ref{eq:mfsol}) that $t$ has
to be close to $t_0$ in order to keep the value of $m(t)$ small, so that the
linearization in (\ref{eq:linear}) is valid.
The magnetization-reversal transition occurs if $m(t_0+\Delta t) \le 0$. 
The phase boundary can therefore be obtained from ({\ref{eq:mfsol}) by putting
$m_w = m(t_0+\Delta t) = 0$. The resulting equation for the phase boundary is then 
\begin{equation}
h_p^c = \frac{m_0 \epsilon T}{1-\exp(-\epsilon \Delta t)}.
\end{equation}
In Fig. 3, this analytic result for the phase boundary has been compared with 
those obtained by solving numerically equation (\ref{eq:mfeq}). As one can 
clearly see from the figure
that the agreement is good in the small $\Delta t$ region of the 
phase boundaries for large values of $T$.

As mentioned before, the magnetization $m_w$ at the 
time of withdrawal of the pulse, seems to be the crucial quantity 
governing the transition.
The sign of \( m_{w} \) solely decides the final equilibrium state of the system 
out of the two equilibrium choices.
Therefore we define the mean field order parameter (\({ \cal O} \)) as following 

\begin{equation}
{\cal O}=m_{w}\theta (m_{w}) ,
\label{eq:mfop}
\end{equation}

\noindent
where the step function $\theta$ is defined as

\begin{eqnarray}
\theta (x) & = & 1, {~~\rm for }~~x>0 \nonumber \\
 & = & 0 , {~~\rm otherwise}. \nonumber
\end{eqnarray}
The nature of variation of ${\cal O}$ with $\mid h_{p}-h_{p}^{c}\mid$ 
at different
values of \( \Delta t \) and $T$ is shown in Fig. 4. Here
$h_p^c=h_p^c(\Delta t, T)$ is obtained from the phase boundary. We fitted 
the order parameter variations to the power law form 
\begin{equation}
 {\cal O} \sim \mid h_{p}-h_{p}^{c}\mid ^{\beta }
\label{eq:mfscal}
\end{equation}
and found the value for the exponent $\beta \simeq 1$ by fitting the 
numerical results for different values of $\Delta t$ and $T$.

The relaxation time (\( \tau  \)) grows as the phase boundary is approached
from either side and shows a divergence on the boundary. We measure the 
relaxation time by measuring the time required by $m(t)$ to reach the final
equilibrium value $\pm m_0$, with an accuracy of $O(10^{-4})$, from the
time of withdrawal of the pulse.
According to (\ref{eq:mfop}) and (\ref{eq:mfscal}), for a fixed $\Delta t$,
$m_w \sim \mid h_p - h_p^c \mid$.
Therefore $\tau$ is also expected to
diverge as $m_w$ vanishes. Fig. 5 shows that this is indeed the case and 
the growth of the relaxation time was found to be logarithmic in
nature. That the relaxation time will diverge logarithmically at the phase
boundary at any temperature below the static critical temperature ($T<1$) can
be shown analytically. Let us follow the mean field dynamics of the system after
the withdrawal of the field. The system starts with a magnetization $m_w$ and 
evolves according to (\ref{eq:mfeq}) with $h(t)=0$ to reach the final state of
equilibrium characterized by magnetization $\pm m_0$. Keeping the value of $m/T$ small
we expand tanh of (\ref{eq:mfeq}) upto the cubic term :
\begin{equation}
\frac{dm}{dt} = \epsilon m + \alpha m^3 ~~,
\label{eq:cube}
\end{equation}
where $\alpha = -1/(3T^3)$.
Thus we can write
\[ \int_0^\tau dt  =  \int_{m_w}^{m_0} \frac{dm}{\epsilon m + \alpha m^3} \]
or,
\begin{equation}
\tau  =  - \frac{\ln m_w}{\epsilon} + \frac{\ln(\epsilon+\alpha m_w^2)}{2\epsilon} + C(T),
\label{eq:tau1}
\end{equation}
where $C(T)$ is a constant depending on temperature only. Now, if the 
combination of $h_p$ and $\Delta t$ is such that one starts with very small 
value of $m_w$, then one gets 
\begin{equation}
\tau \sim - \frac{1}{\epsilon} \ln \mid m_w \mid .
\label{eq:taumw}
\end{equation}
It can be 
easily verified that even if the $O(m^5)$ term in the expansion of 
(\ref{eq:cube}) is kept, the final result (\ref{eq:taumw}) does not
get modified in the $m_w \rightarrow 0$ limit. The above form for the variation 
of $\tau$ fits accurately the numerical results : at $T=0.6$, $1/\epsilon = 1.5$,
and this matches exactly with the numerical result as shown in Fig. 5. The above 
analysis also
makes it very clear that the logarithmic divergence of $m_w$ implies a similar 
divergence with $\mid h_p - h_p^c \mid $ for a fixed $\Delta t$.
It may be mentioned here that for the 
dynamic magnetization-reversal transition, the above divergence in $\tau$ 
occurs at any $T < T_c$. The same equation (\ref{eq:taumw}) also gives the well 
known $\mid T - T_c \mid^{-1}$ divergence of $\tau$ for the static 
transition at $T=T_c=1$. It may also be noted from (\ref{eq:taumw}) that the
logarithmic dependence of $\tau$ on $\mid h_p - h_p^c \mid$, through its 
dependence on $m_w$, for this dynamic transition at $h_p^c(\Delta t, T<T_c)$ is
qualitatively different from the divergence of $\tau$ ($\sim \mid T - T_c \mid^{-1}$)
 for the static transition in the same model (equation \ref{eq:mfeq}) at $T=T_c$.

\section{Monte Carlo Study}

For the MC simulation we have taken Ising spins on a square lattice of size
$L \times L$, 
with $L = 200$ for typical studies. 
Each spin of the lattice were updated sequentially using the Glauber single spin
flip dynamics \cite{bin}. Phase boundaries at different temperatures below the 
static
critical temperature ($T_c \simeq 2.27$) shows qualitatively similar nature to
that in the mean field case. However, unlike the mean field phase diagrams,
the presence of fluctuations causes the MC phase diagrams to touch the 
abscissa asymptotically. When the contributions of fluctuations become 
important at higher values of $T$ and small values of $h_p$,
the above MF theory fails. If $h_p \rightarrow 0$ (as in the large $\Delta t$ 
region of the phase boundary), one can use the picture of nucleation of a 
single domain. The classical nucleation theory of Becker and D\"{o}ring (see 
e.g. \cite{rik2}) 
suggests that the nucleation rate $I \sim \exp[-F(l_c)/T]$ is given by the 
optimality condition $l_c = [\sigma (d-1)/2d]/h_p$ of free energy $F(l) =
2h_pl^d + \sigma l^{d-1}$ for the formation of a droplet or domain of linear
size $l$ under field $h_p$. Here $\sigma$ is proportional to the surface tension
for the formation of a droplet. Equating the growth rate given by Becker and
D\"{o}ring nucleation rate $I$ with the inverse pulse width, one gets $\Delta t \simeq
\exp(1/h_p^{d-1})$, suggesting 
\begin{equation}
h_p^c \ln \Delta t = {\rm constant},
\label{eq:rubbish}
\end{equation}
along the phase
boundary in two dimensions. It agrees fairly well with the MC estimated
phase diagram in the low $h_p$ (i.e. large $\Delta t$) limit \cite{mc,as}.
For large values of $h_p$, the critical droplet size $l_c \sim 1/h_p$ being
much smaller than the system size $L$, many droplets grow simultaneously and the
transition occurs due to coalescence of the droplets and not by the 
above-mentioned 
process of growth of a single droplet. The equation (\ref{eq:rubbish}), therefore, 
fails to fit the phase boundary for large values of $h_p$ and hence for small values
of $\Delta t$.

Similar to the MF case (\ref{eq:mfop}), we define the order parameter in the 
same way : ${\cal O} = m_{w} \theta (m_{w})$, where $\theta$ is the step function 
and $m_w$ is the average magnetization at the time of withdrawal of the pulse.
Fig. 6 shows the variation of this order parameter with $\mid h_{p}-h_{p}^{c}
\mid$ for different
values of $\Delta t$ and $T$. Typical number of initial seed 
values taken to generate the configurations over which each 
data point is averaged is 500. The variation of ${\cal O}$ was again fitted to the power law 
form (\ref{eq:mfscal}) 
with the order parameter exponent $\beta$ and we find $\beta \sim 1$ by fitting 
the MC results for four different values of $\Delta t$ and four different values 
of $T$.

Here also we estimate the relaxation time $\tau$ by measuring the time (MC steps)
required for $m(t)$ to reach the final equilibrium value $\pm m_0$ from the 
time of 
withdrawal of the pulse with a predefined accuracy O($10^{-2}$). Again $\tau$ 
was found to diverge as the phase boundary (at any fixed $T$) is 
approached
from either side. Fig. 7 shows that the divergence of $\tau$ occurs at the point 
where 
$m_w$ vanishes at the phase boundary. It is not
possible to average the relaxation time for different realizations of the
dynamics for a particular $m_w$, as different realizations 
produce different values of $m_w$. Therefore a small range of $m_w$ was taken, 
instead of a particular value, to take an average of $\tau$. A typical 
value of the range of $m_w$ over which averaging has been done is 0.01.
It was observed that $\tau$ depends on the driving parameters $h_p$, 
$\Delta t$ and temperature $T$ only implicitly through the quantity 
$m_w$. The divergence of $\tau$ was found to be very sharp and it fitted
with an exponential function ($\tau \sim \exp \mid m_w \mid $) unlike the mean-field case where $\tau$ was
seen to behave logarithmically with $\mid m_w \mid$.

We fitted above the MC results for $L=200$ and extracted the value of the exponent
$\beta$, assuming the finite size effects to be negligible.
As $h_p \rightarrow h_p^c$, for finite $L$ one should consider the effective 
$h_p^c$ to be a function of $L$. Therefore one should consider an additional 
scaling function of $L/ \xi$ in (\ref{eq:mfop}), where $\xi \sim 
\mid h_p - h_p^c 
\mid^{-\nu}$ is the correlation length with exponent $\nu$. Assuming that $\xi 
\sim L$ for such MC cases, we can write the appropriately modified form of (\ref{eq:mfop}) as
\begin{equation}
{\cal O} \sim L^{-\beta/\nu} f[( h_p - h_p^c ) L^{1/\nu}].
\label{eq:fsc}
\end{equation}
We have fitted the data for different $L$ to the scaling form (\ref{eq:fsc}).
The data for $L$= 50, 100, 200 and 400  for a particular value of $\Delta t$ 
( =5) and two different values of $T$ were scaled to obtain the fitting values 
of the exponents as shown in Fig. 8. 
Typical number of averages over different initial 
seeds taken for the MC data was 2500 for $L=50$ and 50 for $L=400$.
With such a scaling fit procedure, $h_p^c$ could be obtained with an accuracy of
0.001.
This gives $\beta = 0.90 \pm 0.02$ and
$\nu = 1.5 \pm 0.3$. It was observed that the quality of fitting does not
change appreciably for variation of the fitting parameters within the error limits
specified above. 

\section{Discussions}

The recently reported \cite{mc} dynamic magnetization-reversal transition in 
pure Ising model under a pulsed (negative) magnetic
field has been studied extensively in this paper employing the mean field 
approximation as well as the 
Monte Carlo technique in two dimension. Accurate estimates have been made for 
the phase boundary  $h_p^c(\Delta t, T)$, both in MF and MC cases. 
Approximate analytic estimates 
of the phase boundary have been made in both the cases and compared with the 
numerical results obtained. The order parameter for this dynamic
transition has been identified and it is given by the average magnetization at 
the time of withdrawal of the pulse. In the mean field approximation
the order parameter was found to vary linearly with $\mid h_p - h_p^c \mid$,
indicating the order parameter exponent $\beta=1$; while in the MC case $\beta$
was found to be around 0.9 for $d=2$.
The application of the finite size scaling method gives 
the value of the critical exponent $\nu$ for the correlation
length to be nearly 1.5 for $d=2$. Besides the length scale, a time scale was also found
to diverge both in case of mean field and  the Monte Carlo studies. In the mean
field case, the 
relaxation time $\tau$ was found to diverge logarithmically with $m_w$, or 
equivalently with $\mid h_p - h_p^c \mid$, as the phase boundary is approached.
This has been demonstrated to be true analytically also. However, the
divergence of $\tau$ was found to be stronger in the Monte
Carlo case, where $\tau$ diverges exponentially with $\mid m_w \mid$ or with $\mid h_p -
h_p^c \mid^\beta$ (with $\beta \simeq 0.90$ in $d=2$) as the phase boundary is 
approached. The finite size scaling results suggested that the correlation length 
diverges as $\mid h_p - h_p^c \mid^{\nu}$ at the phase boundary with $\nu \simeq
 1.5$ in $d$=2.
The existence
of the divergence of both the length scale and the time scale indicates
the thermodynamic nature of this intriguing dynamic magnetization-reversal 
transition in the Ising model.

\section{Acknowledgements}

We would like to thank M. Acharyya, D. Chowdhury, C. Dasgupta, D. Dhar, A. K. 
Sen and D. Stauffer for some useful comments and suggestions.

\section*{Figure Captions}

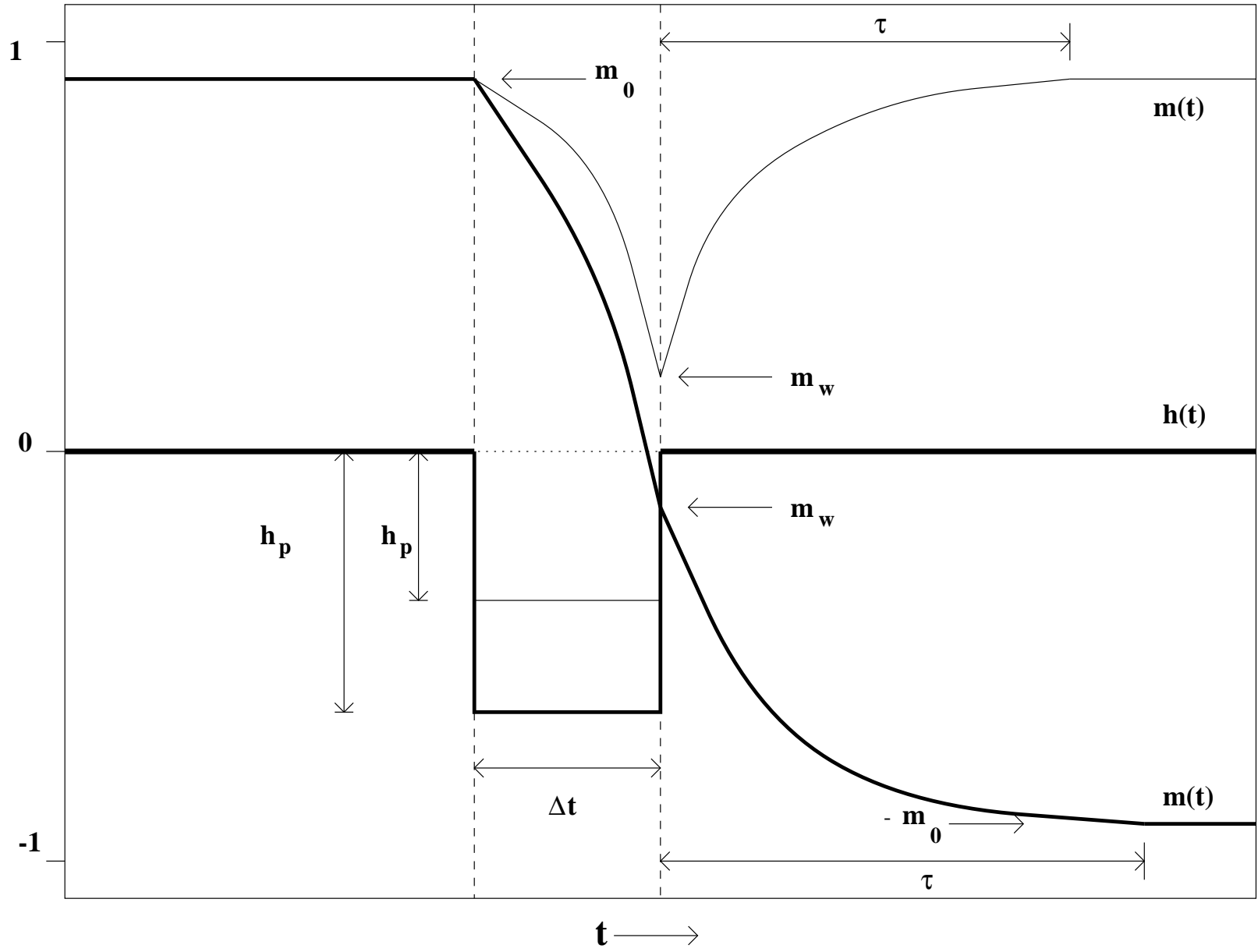
{\bf Figure 1.} Schematic time variation of the external field $h(t)$ and of the 
corresponding response magnetization $m(t)$.
The thin line (smaller $h_p$) shows the case of no-transition, 
whereas the thicker line (larger $h_p$) shows the occurrence of transition. The
various quantities studied in the paper like the relaxation time $\tau$, magnetization
at the time of withdrawal of the field $m_w$ etc are also indicated.

\vspace{0.5cm}
\noindent
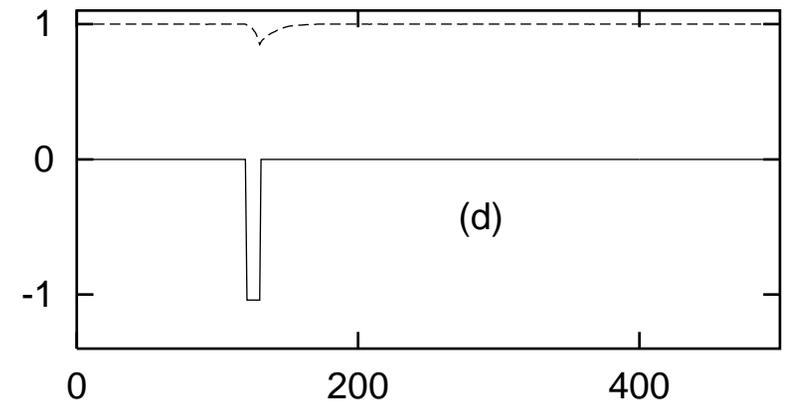
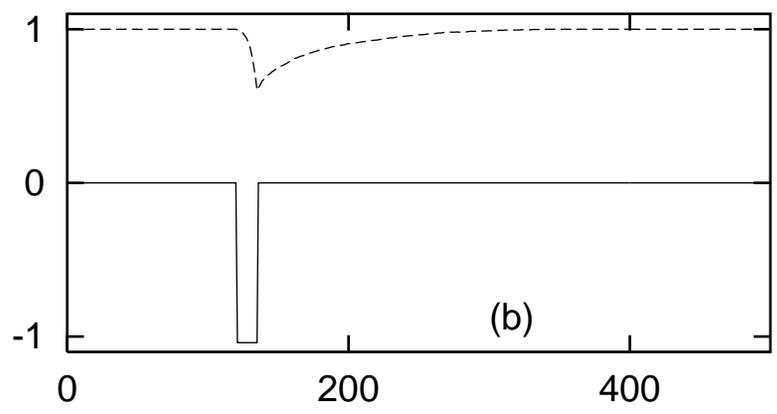
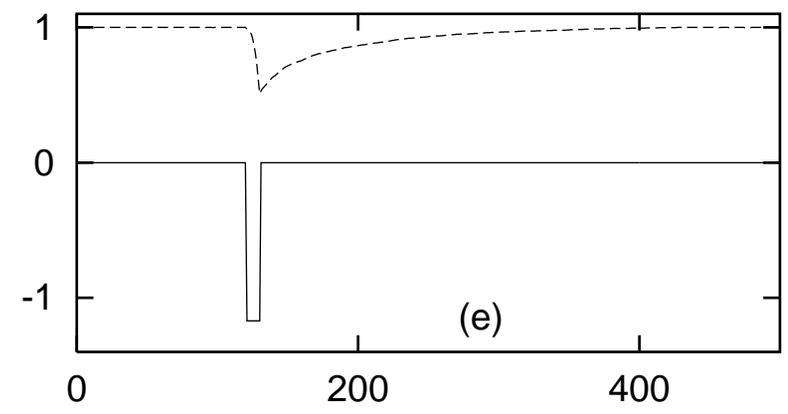
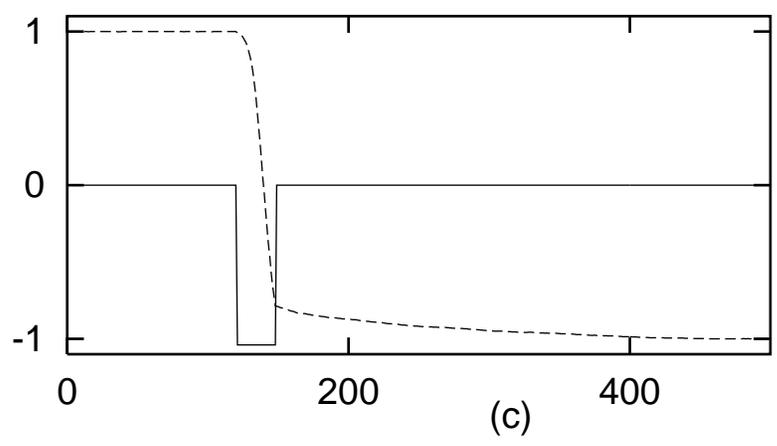
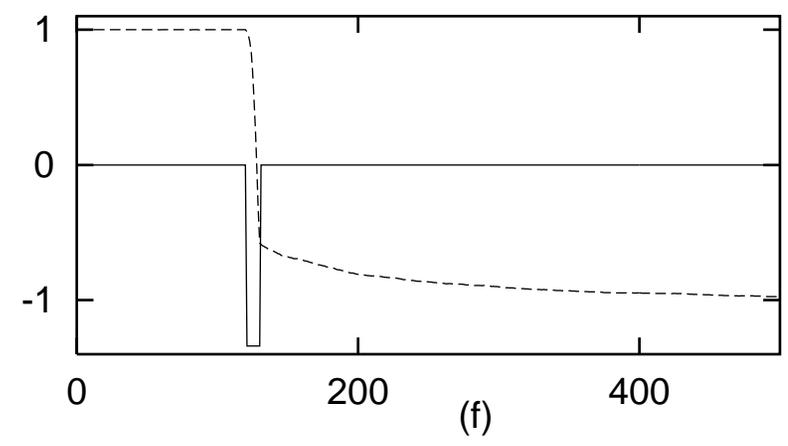
{\bf Figure 2.} Monte Carlo results for the time variation of the magnetization $m(t)$
(dotted line) against the time variation of the field $h(t)$ (solid
line) at $T=1.0$ for (a) $h_p = 1.04$, 
$\Delta t = 10$; (b) $h_p = 1.04$, $\Delta t = 15$; (c) $h_p = 1.04$, $\Delta t = 28$; (d) $h_p = 1.04$,
$\Delta t = 10$; (e) $h_p = 1.17$, $\Delta t = 10$; (f) $h_p = 1.34$, $\Delta t = 10$.

\vspace{0.5cm}
\noindent
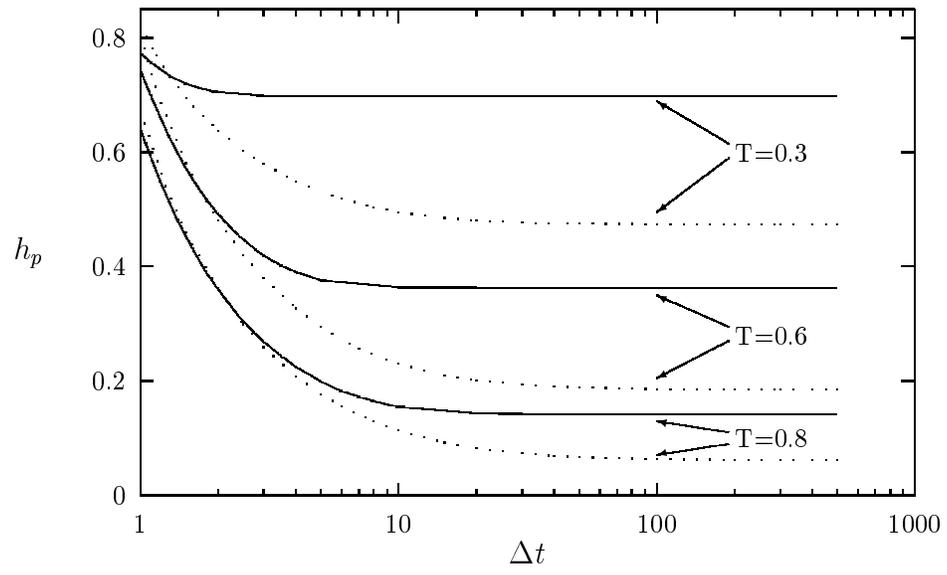
{\bf Figure 3.} Mean field phase boundary in the $h_p-\Delta t$ plane for three
different values of $T$. The dotted lines correspond to the numerical solution of 
equation (\ref{eq:mfeq}) while the solid lines give the corresponding theoretical 
prediction (from equation (\ref{eq:taumw})).

\vspace{0.5cm}
\noindent
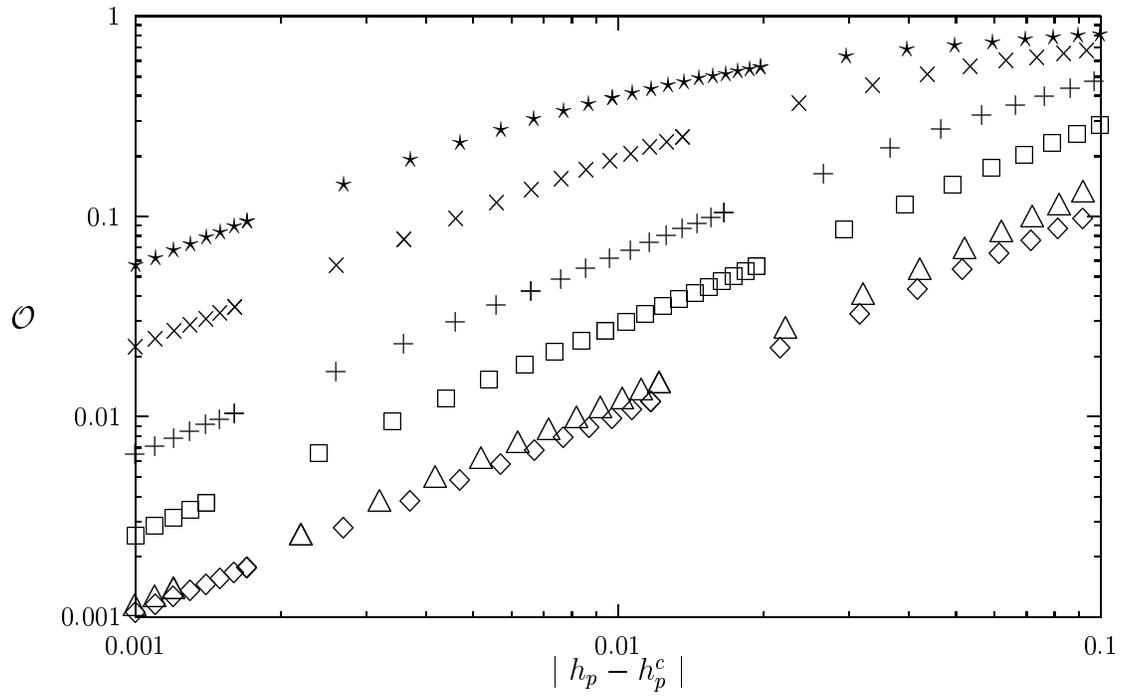
{\bf Figure 4.} Log-log plot of the variation of order parameter ${\cal O}$ against 
$\mid h_p - h_p^c \mid$ for 
$\Delta t=1$, $T=0.9$ ($\Diamond$);
$\Delta t=1$, $T=0.6$ ($\triangle$); 
$\Delta t=2$, $T=0.8$ ($\Box$);
$\Delta t=5$, $T=0.9$ ($+$); 
$\Delta t=10$, $T=0.8$ ($\times$) and 
$\Delta t=10$, $T=0.6$ ($\star$) 
in the mean field case . The $h_p^c$ values are obtained from the corresponding phase 
boundaries (Fig. 3).

\vspace{0.5cm}
\noindent
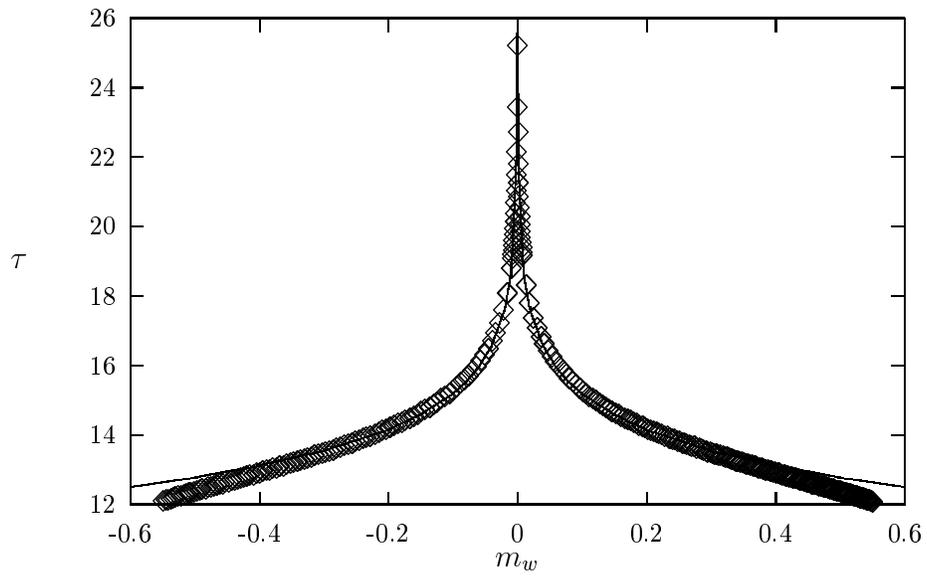
{\bf Figure 5.} Divergence of the relaxation time $\tau$ in MF case. The solid 
line shows the logarithmic fit to the numerical solution of (\ref{eq:mfeq}).

\vspace{0.5cm}
\noindent
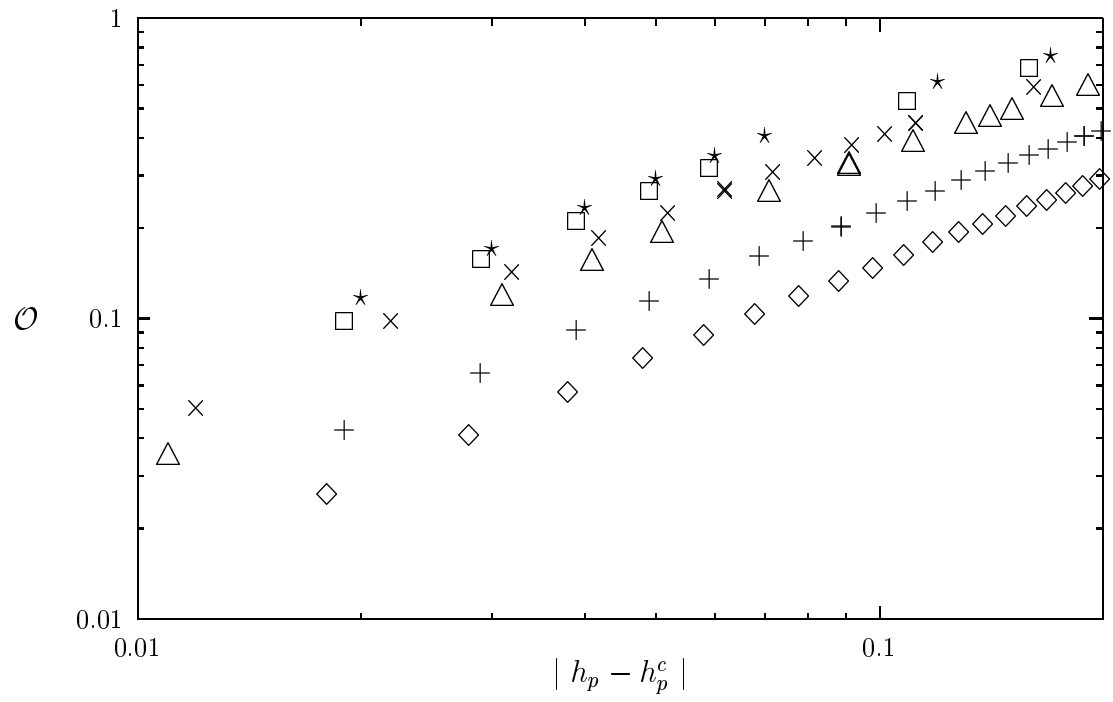
{\bf Figure 6.} Log-log plot of the variation of order parameter {$\cal O$} 
against 
$\mid h_p - h_p^c \mid$ for 
$\Delta t=2$, $T=2.0$ ($\Diamond$);
$\Delta t=1$, $T=1.5$ ($+$);
$\Delta t=5$, $T=1.5$ ($\triangle$);
$\Delta t=10$, $T=2.0$ ($\times$);
$\Delta t=5$, $T=1.0$ ($\Box$) and
$\Delta t=1$, $T=1.0$ ($\star$)
in the MC case. The $h_p^c$
values are obtained from the corresponding phase boundaries in the MC case.

\vspace{0.5cm}
\noindent
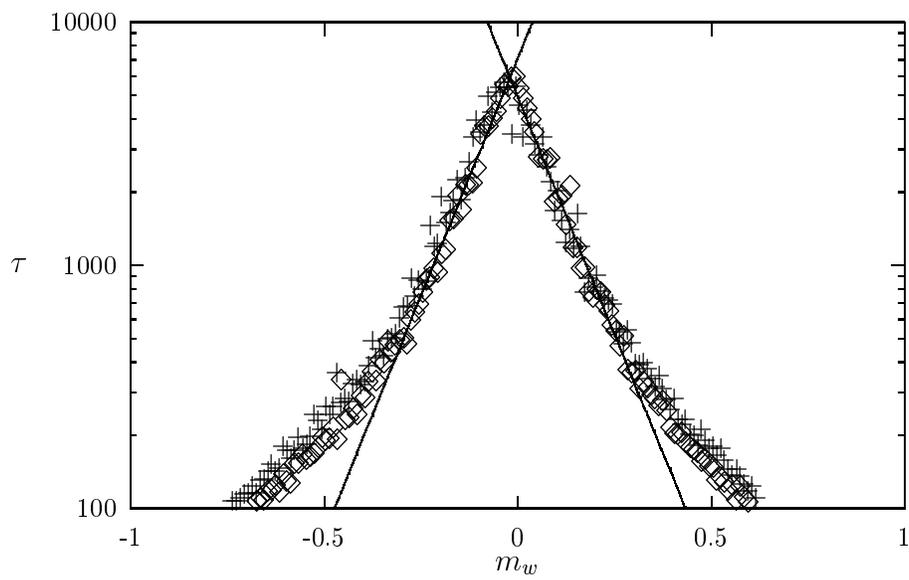
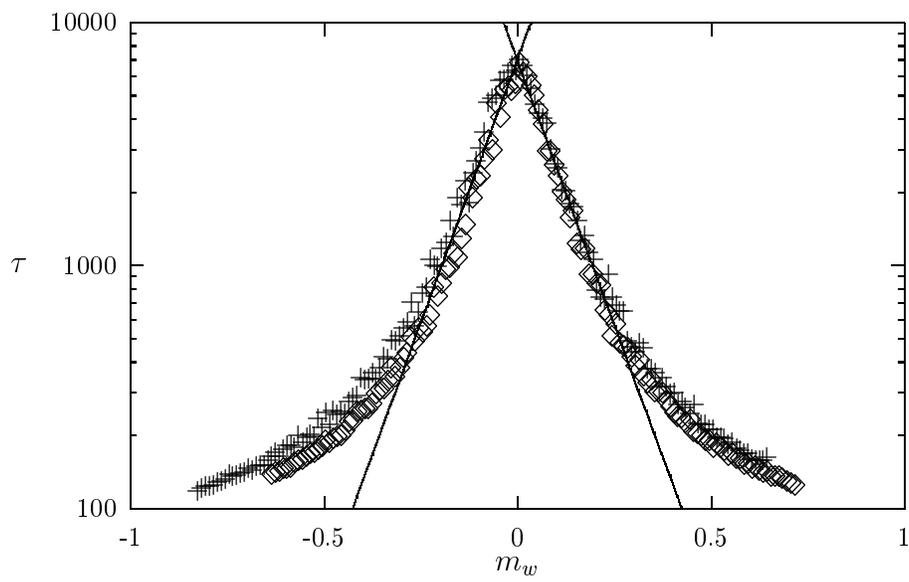
{\bf Figure 7.} Divergence of the relaxation time $\tau$ for
(a) $T=1.5$; $\Delta t=3$ ($\Diamond$), $\Delta t=5$ (+)
and (b) $T=2.0$; $\Delta t=1$ ($\Diamond$), $\Delta t=3$ (+)
in MC case. The solid
lines indicates the exponential fit to the numerical data.

\vspace{0.5cm}
\noindent
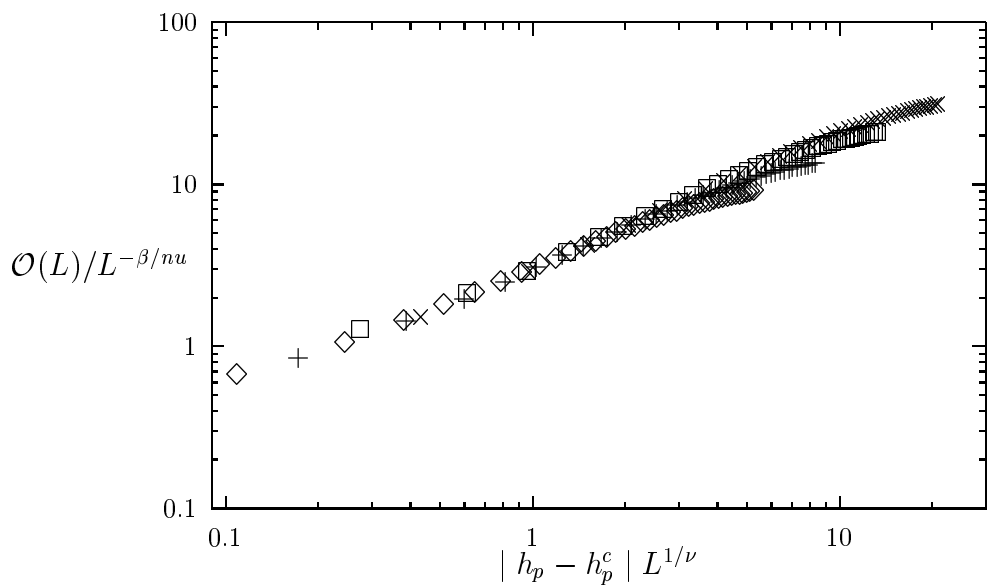
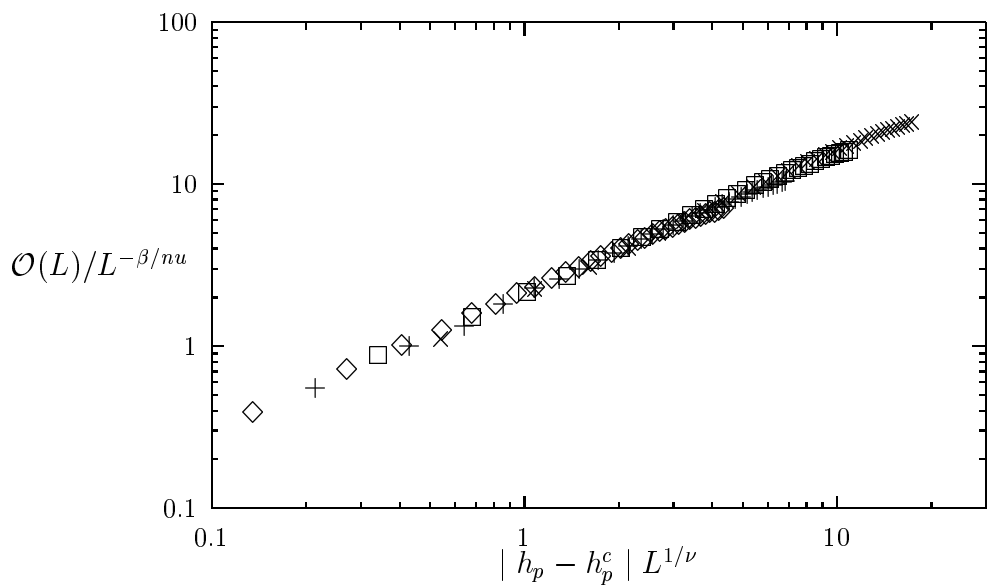
{\bf Figure 8.} Finite size scaling analysis of the MC data: ${\cal O}(L)/L^{\beta/\nu}$ 
plotted against $\mid h_p - h_p^c \mid L^{1/\nu}$ for a fixed $\Delta t$.
(a) $T=1.5$; $L=100 (\Diamond)$, $L=200 (+)$, $L=400 (\Box)$;
(b) $T=2.0$; $L=50 (\Diamond)$, $L=100 (+)$, $L=200 (\Box)$, $L=400 (\times)$.
The best collapsed data are shown
with fitting values $\beta = 0.90$, $\nu=1.5$ and $h_p^c=1.088$ in (a) and $h_p^c=0.720$ in (b).


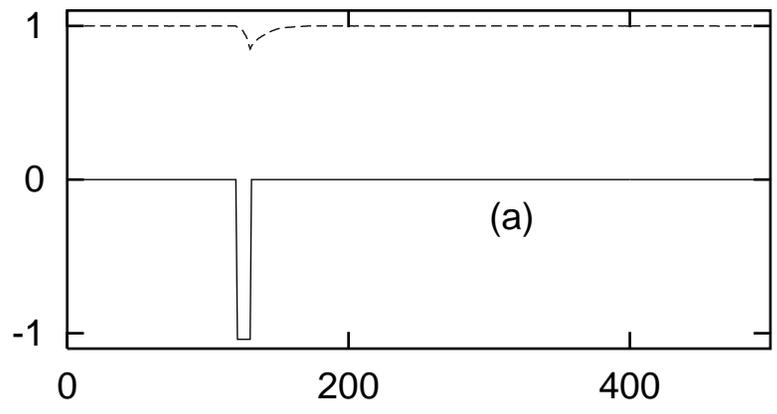
\end{document}